\documentclass[11pt]{article}%
\usepackage[utf8]{inputenc}

\usepackage{amssymb}
\usepackage{amsfonts}
\usepackage{amsmath}
\usepackage{comment}
\usepackage[nohead]{geometry}
\usepackage{setspace}
\usepackage[bottom]{footmisc}
\usepackage{float}
\usepackage{rotating}
\usepackage{gensymb}
\usepackage{lscape}
\usepackage{graphicx}
\usepackage{caption}
\usepackage{subcaption}
\usepackage{siunitx}
\usepackage{todonotes}
\usepackage[small,compact]{titlesec}
\usepackage[hidelinks]{hyperref}
\usepackage{threeparttable}
\usepackage{color}
\usepackage{pbox} 
\usepackage[backend=biber,natbib=true,style=apa]{biblatex} 
\newcommand\numberthis{\addtocounter{equation}{1}\tag{\theequation}}


\title{GPT Adoption and the Impact of Disclosure Policies}

\author{Cathy Yang, David Restrepo Amariles, Leo Allen, Aurore Troussel}

\date{March 1 2025}

\begin{document}
\maketitle

\begin{abstract}
Generative Pre-trained Transformers (GPTs), particularly Large Language Models (LLMs) like ChatGPT, have proven effective in content generation and productivity enhancement. However, legal risks associated with these tools lead to adoption variance and concealment of AI use within organizations. This study examines the impact of disclosure on ChatGPT adoption in legal, audit and advisory roles in consulting firms through the lens of agency theory. We conducted a survey experiment to evaluate agency costs in the context of unregulated corporate use of ChatGPT, with a particular focus on how mandatory disclosure influences information asymmetry and misaligned interests. Our findings indicate that in the absence of corporate regulations, such as an AI policy, firms may incur agency costs, which can hinder the full benefits of GPT adoption. While disclosure policies reduce information asymmetry, they do not significantly lower overall agency costs due to managers undervaluing analysts' contributions with GPT use. Finally, we examine the scope of existing regulations in Europe and the United States regarding disclosure requirements, explore the sharing of risk and responsibility within firms, and analyze how incentive mechanisms promote responsible AI adoption.

\end{abstract}

\subsubsection*{Keywords: Human-GPT collaboration, GPT disclosure, agency theory, content evaluation, survey experiment}

\newpage
\doublespacing

\section{Introduction}

The rapid adoption of generative artificial intelligence (GENAI) technologies, driven by advanced Large Language Models (LLMs) such as GPT, Claude, and Mistral, presents organizations with unprecedented opportunities. However, it also introduces complex legal and risk management challenges that demand careful consideration. A landmark ruling by the Hamburg Labor Court (Arbeitsgericht Hamburg) on January 16, 2024, addressed one of those crucial legal questions: Does an employer engage in employee monitoring when allowing the use of ChatGPT in the workplace via personal accounts? The Court ruled that as long as the employer has no knowledge of which employees are using ChatGPT, when, how, or to what extent, no monitoring occurs  \citep{initial_suit}. This ruling brings attention to a broader challenge, as organizations often struggle to monitor whether and how employees are using generative AI tools. We define the undisclosed use of AI tools in the workplace as ``shadow adoption." To systematically analyze this issue, we propose using agency theory as an analytical framework. 

Despite the expanding body of literature and experiments aimed at understanding the impact of generative AI in the workplace, there is a notable lack of empirical research analyzing this impact through the lens of agency theory. This gap is surprising, as the principal-agent problem can elucidate several critical issues identified in current research. These issues include the motivations and contexts under which employees adopt tools like ChatGPT \citep{DellAcqua2023-mj,brynjolfsson2023generative,peng2023impact}, the conditions that give rise to specific legal risks and strategies for managing these risks \citep{gupta2023chatgpt,hacker2023regulating,charfeddine2024chatgpt}, and the types of monitoring systems and incentives that can be implemented to promote the responsible use of generative AI in firms \citep{humlum2024adoption}.

In this article, we argue that examining the adoption of ChatGPT and similar generative AI software in the workplace through the lens of the principal-agent problem offers valuable insights for developing effective corporate AI policies. Such policies are essential for managing legal risks and promoting responsible AI usage \citep{chiu2021managing}. Our framework specifically highlights the organizational risks associated with individual and team-level adoption of generative AI within professional service firms. By analyzing the principal-agent relationship, we investigate how key factors—namely, information asymmetry, and moral hazard, incentives—shape the risks emerging from the adoption of ChatGPT and generative AI in firms. Accordingly, we later discuss why corporate AI policies are necessary to address these issues and how they can leverage solutions proposed in the agency literature. Specifically, we explore the implementation of control measures, incentive mechanisms, and risk management strategies as effective approaches.

The literature widely acknowledges that an agency relationship is established when a principal hires an agent to perform tasks on the principal's behalf \citep{jensen1976theory,fama1980agency,zeckhauser1985principals}. The principal-agent scholarship focuses on addressing issues that arise from this relationship. These issues typically stem from three fundamental assumptions in agency theory \citep{eisenhardt1989agency,bosse2016agency}. Firstly, issues arise from a misalignment of interests, wherein agents prioritize their self-interest over the objectives of principals \citep{harris1979optimal,bosse2016agency}. Secondly, problems also emerge due to information asymmetry, as agents, who are the primary task performers, possess more information than principals \citep{zeckhauser1985principals,bergen1992agency}. Thirdly, discrepancies in risk preferences between principals and agents lead to divergent actions (moral hazard), with agents generally exhibiting greater risk aversion compared to principals \citep{harris1979optimal,bosse2016agency}. Agency theory posits that effective monitoring and incentive structures are essential for addressing these issues and ensuring risk-sharing that aligns the agent's risk attitudes and behaviors with the principal's interests \citep{jensen1976theory,shavell1979risk,eisenhardt1988agency,miller2005solutions}. While the implementation of monitoring and incentive mechanisms constitutes agency costs for the principal \citep{eisenhardt1989agency}, principals may prefer to incur these expenses rather than face potentially higher costs associated with diminished work quality and inadequate risk management.

Agency theory frequently conceptualizes the relationship between principals and agents using the metaphor of a contract \citep{jensen1976theory,keeley1980organizational,eisenhardt1988agency}. Through this contract, the principal delegates tasks and decision-making authority to the agent \parencite[][p.~308]{jensen1976theory}. This perspective aligns with the modern theory of the firm, which defines firms as a nexus of contracts, where production results from coordinated efforts of a group of individuals \citep{fama1980agency,bratton74thenexus}. Within this nexus, the ultimate principal is the shareholder, and the executives are the agents. Further down the line, managers and employees are linked by principal-agent relationships throughout the production chain \citep{langevoort2003managing}. As Zeckhauser notes, whenever one individual depends on the actions of another to produce value, an agency relationship arises: ``The consultant is the agent, the client the principal. The absence of a formal contractual agreement directly binding managers and employees does not impede the establishment of a principal-agent relationship \citep{spremann1987agent}, as they are functionally interconnected and linked through the nexus of contracts constituting the firm. In other words, the contracting relationship between principal and agent may be either implicit or explicit \parencite[][p.~308]{jensen1976theory}.

The chain of agency relationships within the corporation has been extensively discussed in legal scholarship \citep{easterbrook1989corporate,brudney1996contract,langevoort2003managing}. Langevoort provides a thorough examination of what he terms ``agency law within the corporation" to elucidate the duties of employees throughout the organization through the lens of principal-agent relationships. He highlights the intra-corporate duty of candor, which defines the duty owed by the board of directors to the shareholders, as well as by subordinates to their superiors \citep{langevoort2003managing}. According to Langevoort, this duty is essential for ensuring that critical information flows upwards to the board of directors, enabling them to be accountable to the shareholders. Consequently, corporate officers and employees are bound by an obligation that extends beyond merely avoiding falsehoods; they have an affirmative duty to disclose pertinent information. This obligation includes delivering all information relevant to the superior's decision-making, including any risks known to the employee \parencite[][p.~2111]{langevoort2003managing}.

Relationships where firms hire professional service organizations, such as consulting, auditing, and law firms, to carry out specific technical tasks are well-documented as principal-agent exchanges \citep{sharma1997professional,eisenhardt1989agency,mills1990agency}. However, this article focuses on the internal dynamics between employees within the same professional service firm, particularly consulting firms tasked with delivering services to external clients, including strategic business advice, legal support, and auditing services. In this context, the agency relationship is established between a principal—an employee of the consulting firm who defines and assigns tasks and decisions—and an agent, another employee hired by the consulting firm to execute these tasks under the principal's direction.

One distinctive feature of agency relationships in consulting firms is the involvement of knowledge workers \citep{sharma1997professional,mitchell2011knowledge}. Unlike machine work, the success of knowledge-based organizations hinges on the abilities and expertise of their employees, who generate valuable ideas and knowledge that are not always easy to control. For example, managers may delegate specialized tasks to team members, such as financial, legal, or data analysis. Team members, acting as agents, use their knowledge and skills to complete these tasks, producing outputs that heavily depend on their specific competencies. As Mitchell and Meacheam state, following \textcite[][p.~759]{sharma1997professional} ``the most significant difference between traditional agency exchanges and those involving knowledge workers is that, where a knowledge worker is the agent, division of labor is also a division of knowledge" \parencite[][p.~151]{mitchell2011knowledge}. Characterizing the principal-agent relationship in consulting firms as a knowledge agency relationship highlights the specific knowledge effort and input that agents are expected to provide to the principal. This perspective anticipates the role that ChatGPT and similar technologies can play in exacerbating some of the principal-agent problems.

In this article, we specifically analyze the dynamics between managers in legal, audit, and advisory roles, and their subordinates, which include both junior lawyers, analysts, and consultants, particularly focusing on scenarios where these subordinates utilize ChatGPT to perform their tasks. We conceptualize this dynamic as the manager-agent relationship. Within this framework, subordinates act as agents under the supervision of managers, who serve as principals, with the goal of delivering services to the firm's partners and clients. By interpreting the manager-subordinate relationship through the lens of agency theory, we scrutinize the operational and legal risks created by the agents' adoption of ChatGPT, focusing on information asymmetry and moral hazard. Additionally, we examine why AI corporate policies are essential for effectively addressing these novel risks and identify the specific issues these policies should consider.

First, the agent's use of ChatGPT may exacerbate information asymmetry between the principal and agent, particularly when the manager is unaware of the agent's reliance on ChatGPT. Without this knowledge, managers cannot implement additional controls to ensure the quality, authenticity, and originality of the work produced. The management literature and this study consistently show that managers have a hard time distinguishing between work produced by consultants with and without the assistance of ChatGPT due to the human-like text generated from ChatGPT \citep{dou2021gpt,crothers2023machine}. Consequently, this new form of information asymmetry impedes managers' ability to assess well-established risks associated with ChatGPT, such as confidentiality breaches and misinformation. Finally, managers also lack information about the extent and types of interactions between the agent and ChatGPT, making it difficult to monitor the effort produced by the agent and the quality of the production process, let alone the process and quality of the outputs of ChatGPT itself. 

Second, agents using ChatGPT may engage in behaviors that are not in the principal's best interest, thereby introducing moral hazard. Moral hazard arises because agents may prefer to exert less effort if all other factors remain constant \citep{harris1979optimal}. While the principal might be indifferent to the agent's effort level if their share of benefits remains unchanged, they may nonetheless be exposed to additional, unrecognized risks. For instance, agents might opportunistically use ChatGPT to complete tasks quickly without substantial personal effort, given that managers still perceive the output quality favorably. Managers, unaware of the extent of ChatGPT usage, might erroneously assume that analysts are dedicating significant time and effort to their tasks, and pass on wrong information on costs to the partners and the client. The agent's risk preferences also differ from those of the principal. Agents may be more inclined to take risks, such as sharing confidential information with ChatGPT or incorporating ChatGPT-generated content into deliverables, if the principal is unaware of its use and cannot distinguish the source of the output. This misalignment in risk preferences and information can lead to suboptimal outcomes for the firm. Specifically, while consultants may benefit from the efficiency gains of using ChatGPT, the firm bears the risks associated with misinformation and confidentiality breaches.

The potential for high-risk exposure and monitoring costs could prevent managers in consulting firms from adopting enterprise-wide GPT despite its benefits. This conjecture is consistent with recent literature indicating that individuals' low adoption rate of LLM in the workplace is primarily attributed to restrictions on use from their superiors \citep{humlum2024adoption}, which highlights that corporate adoption of GPT largely relies on the incentives of the principal (manager) among the principal-agent dyad to adopt GPT that aligns with the interests of the agents (analysts). 

Traditional technology acceptance model suggests that technology adoption is positively driven by potential users' perceived usefulness and negatively influenced by perceived risk associated with the technology \citep[e.g.,][]{agarwal1998conceptual, davis1989perceived, venkatesh2000theoretical}. Despite the perceived usefulness with manageable risks under the recent technological advancements in AI, extensive literature documents individuals' aversion to its adoption \citep{dietvorst2015algorithm, longoni2019resistance}, especially when the applied field is more subjective than objective \citep{castelo2019let}, or when the individual is more experienced with the task \citep[e.g.,][]{logg2019algorithm, DellAcqua2023-mj}. Enhancing algorithmic transparency is one critical method to promote AI adoption. Transparency in AI is defined by disclosure in its usage, data handling, and decision-making processes \citep{lage2019human,poursabzi2021manipulating}. While existing empirical studies show that increased algorithmic transparency could encourage adoption by enhanced decision accountability \citep{Gregor.1999}, trust \citep{Wang.2007}, perceived fairness \citep{dodge2019explaining}, and task efficiency \citep{senoner2021using}, other studies find the opposite due to information overload \citep{poursabzi2021manipulating, you2022algorithmic} and reduced trust \citep{dietvorst2015algorithm}. It is, therefore, unclear whether increasing transparency via subordinates' disclosure of the GPT use could encourage managers' adoption of GPT, which we empirically investigate in this study. 

Prior research about GPT adoption focuses on individual-level adoption without considering the corporate hierarchies \citep{Noy2023-se, DellAcqua2023-mj}. Yet, discussing the superiors' (managers') incentive to adopt GPT is challenging without considering implications for the subordinates contracted to perform specific tasks. When managers act as principals in a principal-agent dyad, subordinates' disclosure of GPT use implies reduced information asymmetry. Managers incur less information disadvantage regarding whether GPT participated in content generation. While this reduced information asymmetry could imply incentives for managers to allow GPT use at the workplace, it could challenge the current contract enforcing knowledge exchange between superiors and subordinates and thus incur agency costs. First, managers allowing GPT usage accompanied by subordinates' disclosure could imply misattribution of the effort subordinates exerted on a task, given its outcome with GPT participation. Thus, this disincentivizes subordinates to adopt GPT or disclose GPT use. Second, disclosure of GPT use increases the subordinates'  accountability in managing risk in content production \citep{fama1980agency,rubin2016principal}, which entails misinformation produced by GPT because of hallucinations and customers' confidentiality breaches due to subordinates' personal GPT usage. In sum, subordinates' disclosure of GPT reduces information asymmetry in a principal-agent dyad. Still, it is unclear whether it would better align the incentives between the manager and their subordinates in a workplace where the availability of GPT disrupts contract enforcement through increased content risk and uncertainties in effort exerted by subordinates, which calls for an empirical investigation.  

In this study, we aim to empirically investigate whether subordinates' disclosure could encourage managers to adopt GPT (i.e., allowing subordinates to use GPT for content generation) in a context where there is no explicit regulation in the corporate AI policies. Additionally, we investigate whether analysts' disclosure of GPT use changes the incentive alignment of the manager-analyst dyad. In particular, we try to provide a nuanced understanding of the trade-offs involved in disclosure practices and their effectiveness in mitigating risks related to misinformation, confidentiality breaches, and the overall quality of work produced by ChatGPT agents. By exploring these dynamics, we contribute valuable insights into designing and implementing corporate AI policies that can better manage emerging risks preventing the adoption of generative AI in professional service firms.

\section{Survey Experiment and Data}
\subsection{Study Overview and Design} 

Our objective is to explore whether analysts' disclosure of the GPT use that potentially reduces information asymmetry and mitigates risk could encourage managers' adoption of GPT in a consulting firm. There are several challenges to exploring this issue causally. First, we must collaborate with a consulting firm that hasn't adopted the GPT for the enterprise because any existing corporate-wide adoption could distort managers' preference for adoption. Second, to allow perfect counterfactual, we also need to enable some managers within a consulting firm to be randomly exposed to the disclosure of GPT use and some not by minimizing the spillover of treatment. Third, the analysts' choice to disclose GPT use needs to be out of the control of the managers to avoid endogeneity introduced by analysts' self-selection and their expectations of how managers react to GPT usage. Fourth, researchers also need to know whether analysts actually used GPT in content generation, which could be close to impossible due to the human-like nature of the GPT output, yet crucial to understanding the extent of information asymmetry induced by GPT usage. Following this point, managers also need to evaluate content generated by analysts with and without GPT to understand the extent of information asymmetry on an individual level. These challenges make achieving our core research objective in a natural field setting difficult, if not impossible. 

To address these challenges, we collaborated with a major consulting firm that regulates the use of GPT at work (e.g., prohibiting employees from including their clients' information in prompts) to conduct a survey experiment. The firm's mid-level managers in legal, audit and advisory roles were enrolled in a training session to help them understand how they can benefit from generative AI to provide data-driven solutions for their potential clients. These mid-level managers often create pitch content with the help of their analysts in response to request-for-proposals (RFPs), which call for projects from potential new clients \citep{falkner2019identifying}. We simulate a critical part of this content co-creation process where managers contract analysts to draft initial pitch content, a research brief, in response to an RFP without knowing the actual content generation process unless disclosed. The experimental design randomly assigns managers to different rooms during their training sessions to evaluate the research briefs drafted by the analysts. One room of managers is assigned a survey link containing the pitch content without disclosing the source, while the other room is assigned a different survey link that discloses the content generation source (for both no-GPT and human-GPT generated content). Managers' physical separation would enforce the disclosure variation in the survey content to prevent treatment contamination due to the potential leak of deck generation source information if the managers were seated in the same classroom. 

To further enforce the analysts' choice to disclose GPT out of the control of the managers, we hired two master's students with previous experience in consulting from a top European business school as junior analysts. The two analysts had no prior interaction with the managers, so they could create the content without knowing the managers' preference for content generation with or without GPT. We developed four RFPs with fictitious client names to increase the generalizability of our findings regarding the business topic in the interest of the consulting firm. Following a general structure of research brief in response to RFPs, each research brief comprised six sections: a problem statement, solution overview, implementation details, expected outcomes, team, and timeline. We obtained the ground truth of the content generation source by first instructing the two analysts to create research briefs collaboratively without using any LLMs, which they named the ``No-GPT deck." After completing the four No-GPT decks, we asked the analysts to use ChatGPT 3.5, the version of the GPT model available for personal use at the time of the experiment, to modify the No-GPT decks without any specific instructions on how to prompt ChatGPT, which they named ``Human-GPT deck."\footnote{Note that generating human-GPT collaborative content requires a longer time than the human-generated content by our design because the total effort amounts to the number of hours analysts spent building the non-GPT deck plus the interactions with GPT. This suggests an underestimation of effort exerted in the Human-GPT deck if managers perceived equal or less time in the content generation by analysts using GPT versus not.} It took the two analysts 38.5 hours with 17€ per hour wage to finish generating the eight decks (four No-GPT decks and four Human-GPT decks), ranging from six to 11 pages each. 

\begin{figure}
    \caption{Type of prompt used in each content generation section}
    \label{fig:prompt}
    \centering
    \begin{subfigure}[b]{.5\textwidth}
        \centering
        \includegraphics[width=.9\textwidth]{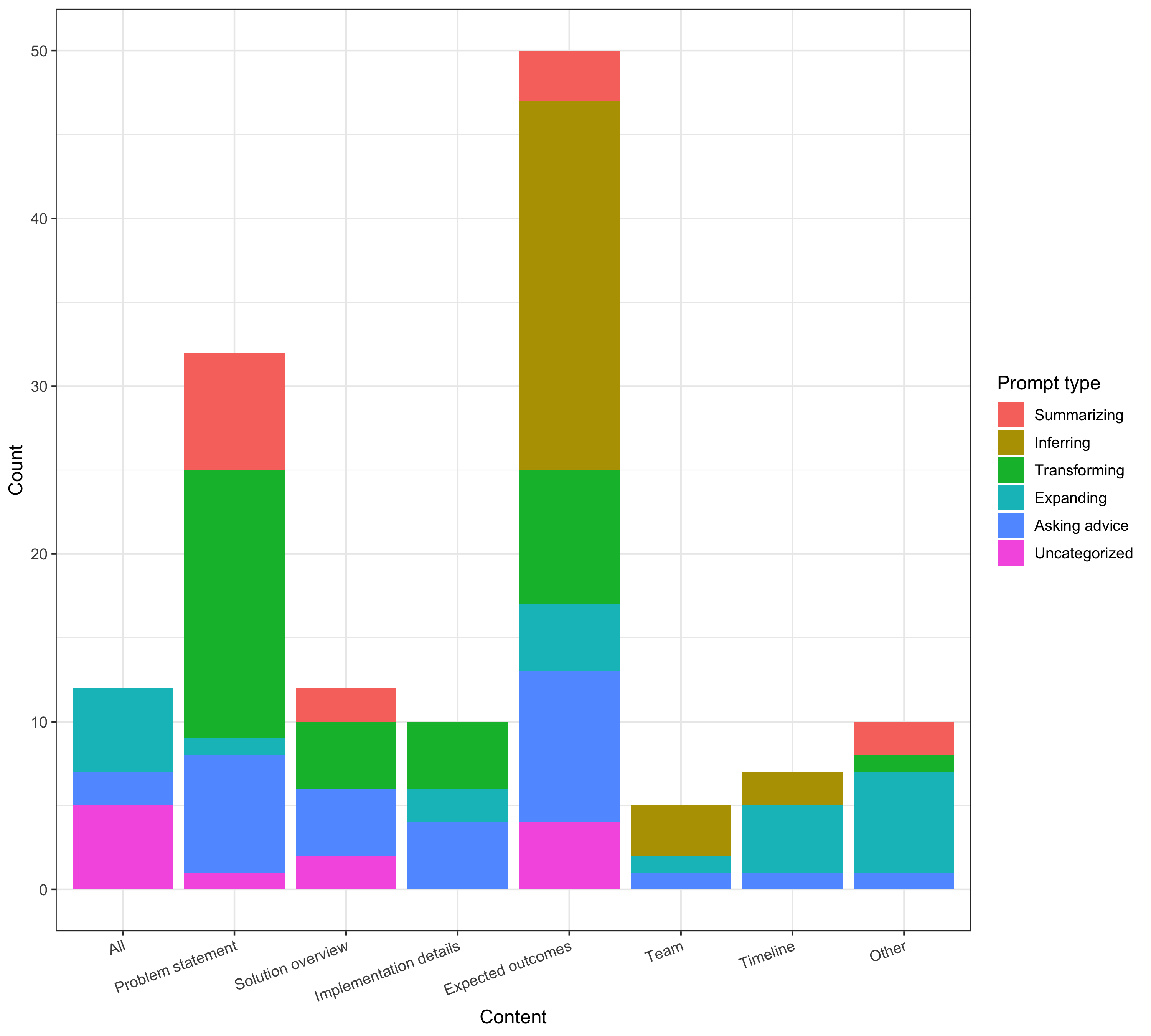}
        \caption{\label{fig:prompt_count}Count}
    \end{subfigure}\hfill
    \begin{subfigure}[b]{.5\textwidth}
        \centering
        \includegraphics[width=.9\textwidth]{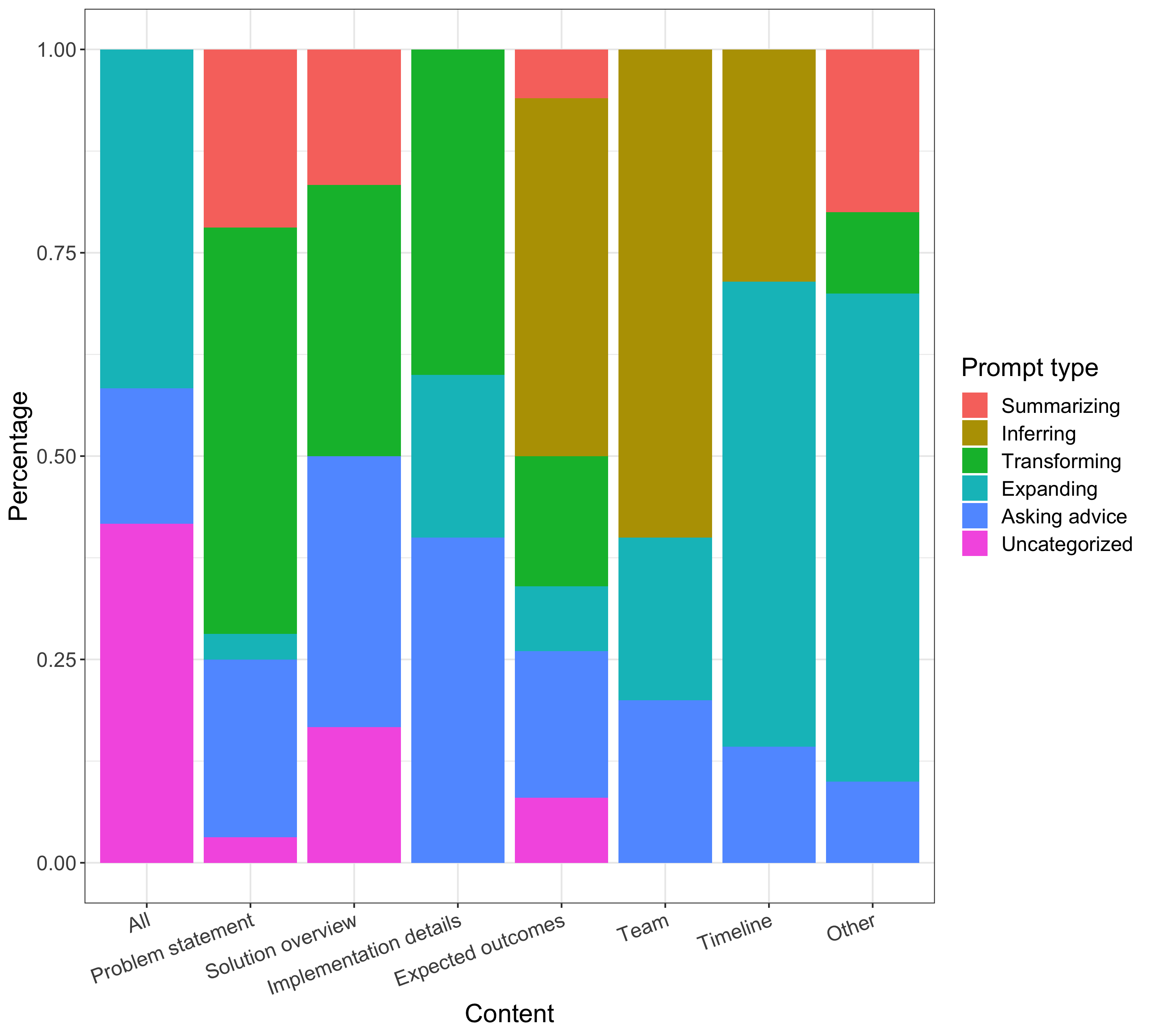}
        \caption{\label{fig:prompt_percentage}Percentage}
    \end{subfigure}
    \vspace{0.5em}
    \begin{minipage}{0.85\textwidth}
    \raggedright\small{\textit{Notes}: We classify the content category of a prompt in following categories: a whole (labeled as ``all") when analysts demanded general assistance, one of the six sections (problem statement,  solution overview, implementation details, expected outcomes, team, and timeline), or other content such as titles.}
    \end{minipage}
    
\end{figure}

To ensure that GPT contributed to the content difference between the No-GPT and Human-GPT deck, we requested the analysts document their prompts and GPT responses by exporting their interactions with ChatGPT. The two analysts used 138 prompts before finalizing the four Human-GPT decks. Figure \ref{fig:num_prompts} shows the number of prompts used to update the content across the six content sections (problem statement, solution overview, implementation details, expected outcomes, team, and timeline). The prompt details are shown in Figure \ref{fig:prompt}. We categorized the prompts into four categories based on their application purpose: summarizing, expanding, inferring, and transforming according to \cite{Ng2023}, with additional categories of general advice-seeking and `uncategorized' when the purpose of the prompt is hard to identify. 

\begin{figure}[!h]
    \centering
    \caption{\label{fig:gpt_contribution}The degree of GPT contribution to the Human-GPT deck}
    \includegraphics[width=0.5\linewidth]{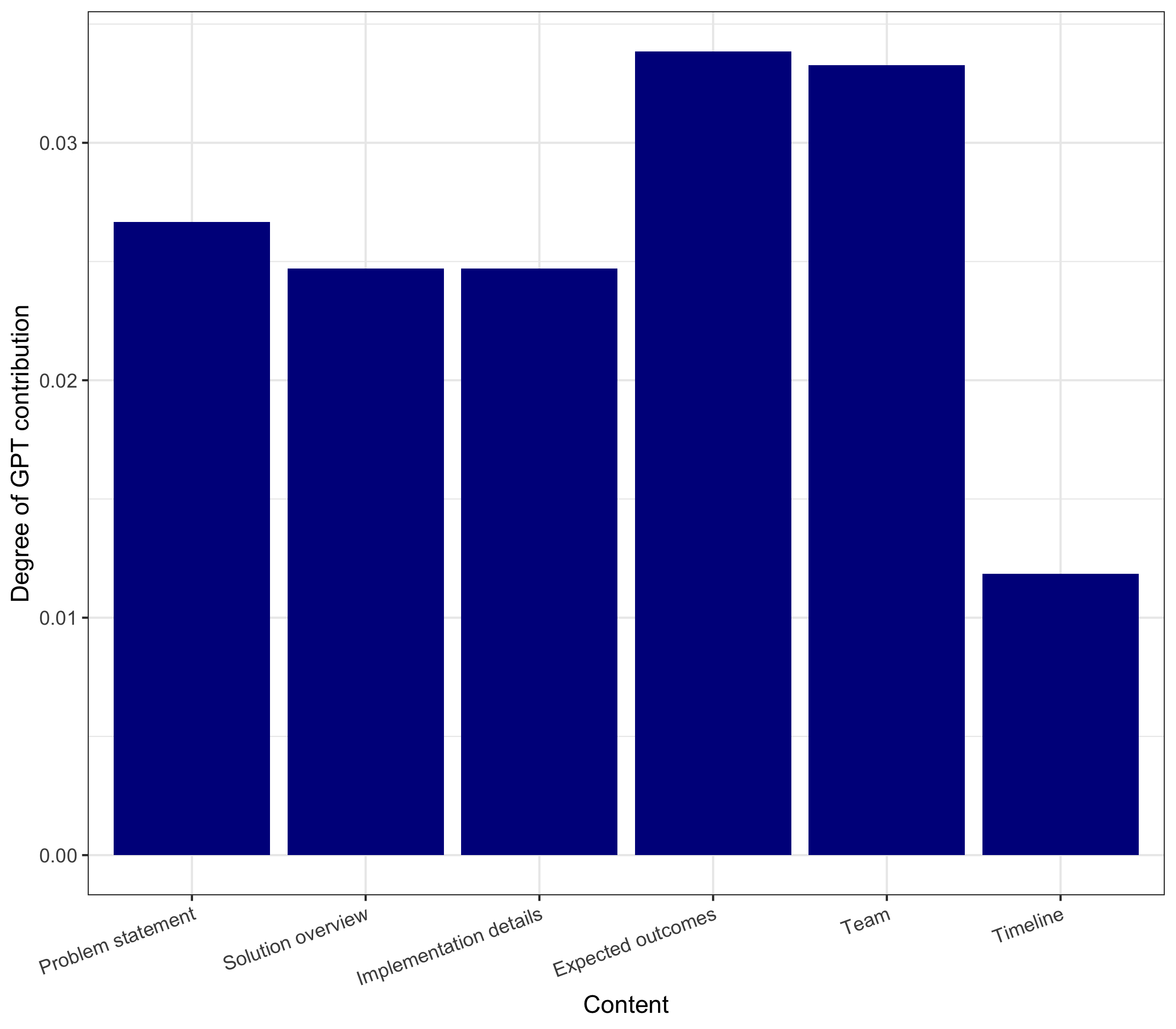}
\end{figure}

We identify the contribution of GPT to the Human-GPT deck by the absolute differences in content similarity between the two versions of the brief (No-GPT and Human-GPT decks) and GPT responses. The content similarity between a deck and GPT responses is defined by the cosine similarity based on the text tokens using BERT tokenization ranging from 0 to 1, where a higher number indicates higher content similarity \citep{devlin2018bert}. Figure \ref{fig:gpt_contribution} shows the degree of GPT contribution to the content in the Human-GPT (versus the No-GPT) deck, which suggests that the Human-GPT deck content results from analysts incorporating the GPT responses.  

After controlling the experimental environment and content generation process, we aim to test empirically: 1) whether there's an information asymmetry that managers cannot identify analysts' use of GPT in generating the research brief without an enforced disclosure policy, and whether GPT disclosure could reduce such information asymmetry; 2) the incentives (mis)alignment in GPT adoption and its disclosure among the manager-agent dyad. We thus adopt a mixed experimental design that combines within-subject and between-subject designs. The between-subject design focuses on the disclosure of content generation sources (no-GPT or human-GPT), compared to no such disclosure. The within-subject design involves each manager evaluating the No-GPT and Human-GPT versions of the research brief for a specific RFP. To identify the potential information asymmetry of analysts' GPT use, we measure managers' belief in analysts' GPT use before and after evaluating a research brief generated with or without GPT under different experimental conditions varying in its disclosure policy. Additionally, we assess analysts' incentives (disincentives) to adopt GPT and disclose its use through managers' evaluations of research brief quality (performance) and perceived effort exerted with and without GPT use \citep{rubin2016principal}. In particular, several scenarios exist: 1) analysts have a greater incentive to adopt and disclose GPT use when managers consider the research brief to be of better quality, and this much-improved quality is attributed to their more significant effort that translates to higher wages; 2) analysts have a greater incentive to adopt without disclose GPT use (i.e., shadow adoption)  when managers consider the research brief to be of better quality, but the improved quality is not attributed to their effort; 3) analysts have no incentive to adopt GPT if the managers do not perceive a quality gain after using GPT for content generation. We define incentive alignment following a disclosure policy when it simultaneously encourages GPT adoption by the manager and analyst. 

The detailed experimental procedure is described in the following section.

\subsection{Experimental Procedure} 

One hundred thirty mid-level managers from a major consulting firm were involved in the training program about using generative AI to provide data-driven solutions across two sessions. All managers were randomly assigned to two classrooms for a training session (90 minutes and 60 minutes in the first and second waves, respectively). Before starting the training session, all managers were told about their mission to evaluate the research brief generated by two junior analysts in response to RFP from potential clients anonymously. In the first round of data collection, two out of four RFPs were randomly selected for each manager; in the second round, one out of four RFPs was chosen randomly. 

Before proceeding to the deck evaluation section, we asked all managers for their estimates on the percentage likelihood that the two junior analysts would use GPT to develop research briefs. Understanding these prior beliefs helps eliminate potential confounding factors regarding how differences in those initial beliefs affect their post-belief perceptions of content generation sources—following disclosure or lack thereof. 

In the deck evaluation section, managers assessed the No-GPT and Human-GPT decks created in response to an RFP, presented in a randomized order. In one classroom, managers were informed about the source of the content generation for the deck (Non-GPT and Human-GPT), while those in another classroom did not receive this information. For each RFP research brief deck, managers rated the content quality using three items on a scale from one to seven, as shown in Table \ref{tab:item} (Cronbach alpha = 0.78). Furthermore, we requested that the managers estimate how many hours they believed the deck generation required from the two analysts to reflect their perceived effort exerted by the analysts. 

Following the presentation of the two decks for each research brief, managers were requested to express their beliefs in the junior analysts utilizing GPT for the No-GPT and Human-GPT decks. We asked managers to indicate their familiarity with the RFP process on a scale from one to five, as previous research shows that experience significantly influences an individual's appreciation or aversion to adopt an algorithm \citep{logg2019algorithm, DellAcqua2023-mj}. 

Following the evaluation of the deck, we assessed managers' tendencies to adopt GPT by their views on the GPT's perceived usefulness and risks \citep{agarwal1998conceptual,davis1989perceived,venkatesh2000theoretical}. We asked participants to rate the perceived usefulness (three items on a scale from one to seven as shown in Table \ref{tab:item}, with a Cronbach alpha of 0.78) and the perceived risk (two items on a scale from one to five as displayed in Table \ref{tab:item}, with a Pearson correlation of 0.43) of using GPT to create a research brief. Managers also indicate their preference to authorize junior analysts to use GPT when preparing a research brief for each of the six content sections (i.e., the problem statement, solution overview, implementation details, expected outcomes, team, and timeline) on a scale of one to five. We allow heterogeneity in managers' GPT adoption preferences across different content sections due to variations in how analysts use GPT, as shown in Figures \ref{fig:num_prompts} and \ref{fig:gpt_contribution}. 

Before proceeding to the end of the survey, the managers were also asked to provide their demographic information, including gender, whether their role in the consulting firm involved facing the clients directly, tenure in the consulting industry, and average weekly working hours (less than 45 hours, between 45 and 50 hours, between 50 and 60 hours, greater than 60 hours). 

\begin{figure}[htbp]
    \centering
    \caption{Experimental procedure}
    \label{fig:experiment_method}    
    \includegraphics[width=0.4\linewidth]{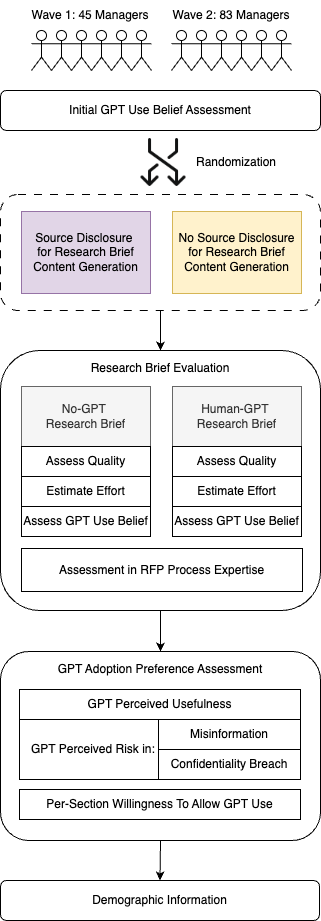}
\end{figure}

Figure \ref{fig:experiment_method} summarizes the experimental procedure.

\subsection{Data and Randomization Check}

Ninety-two managers completed the survey experiment successfully. Among the 92 managers, 43 were informed about the source of the content generation, while 49 were unaware. We conducted a binomial test on the between-subject assignment outcome and found no significant difference in the frequency of managers assigned to the source transparency condition with 0.5 ($p = 0.47$). 

Table \ref{tab:sumstat} displays the summary statistics of the variables used in the main analysis, with the explanations of the variables shown in Table \ref{tab:varname}.

\begin{table}[h]
    \centering
    \caption{Summary statistics\label{tab:sumstat}}
    {
\def\sym#1{\ifmmode^{#1}\else\(^{#1}\)\fi}
\begin{tabular}{l*{1}{ccccc}}
\hline\hline
                &     Mean&  Std Dev&      Min&      Max&Observations\\
\hline
SecGPTAuth      &    2.518&     1.29&        1&        5&      552\\
DeckQuality     &    4.478&     1.02&     1.50&        7&      238\\
DeckHour&    15.77&     27.4&        0&      200&      238\\
IndNoGPTDeckFirst       &    0.511&     0.50&        0&        1&      92\\
IndExperience &    1.672&     0.95&        1&        4&      92\\
IndGPTUsefulness&    5.082&     1.31&        1&        7&       92\\
IndGPTRisk      &    4.571&     0.69&        2&        5&       92\\
IndFemale       &    0.391&     0.49&        0&        1&       92\\
IndClientFacing   &    0.696&     0.46&        0&        1&       92\\
\# Prompts   &    4.83&     6.06&        1&        27&       24\\
\hline\hline
\end{tabular}
}

\end{table}

We conduct a randomization check on the individual-level variables. First, we find no statistically significant difference in managers' \textit{prior} beliefs about analysts' GPT use before evaluating the decks ($M_{No-Disclosure} = 44.9$, $M_{Disclosure} = 44.9$, $p=1.00$ given a two-sample Student's \textit{t}-test). 
There is no statistically significant difference in the proportions of managers with a client-facing role, tenure in the consulting industry, and average weekly working hours between the disclosure and no disclosure conditions ($ps>0.10$ given a Chi-square test of independence). We find proportionately more female managers with marginal statistical significance ($M_{No-Disclosure} = 30.61\%$, $M_{Disclosure} = 48.84\%$, $p=0.07$ given a Chi-square test of independence) in the disclosure (vs. no-disclosure) condition, which we control for in the analyses.

\section{Main Results}\label{main_result}
In this section, we conduct an empirical test based on our survey experiment to understand: 1) whether there's an information asymmetry that managers cannot identify analysts' use of GPT in generating the research brief without an enforced disclosure policy, and whether GPT disclosure could reduce such information asymmetry; 2) the incentives (mis)alignment in GPT adoption and its disclosure among the manager-agent dyad.

\subsection{Impact of the Disclosure Policy on Information Asymmetry in Analysts' GPT Adoption}

\begin{figure}[htbp]
    \centering
    \caption{Managers could not identify analyst's use of GPT unless disclosed.\label{fig:post_belief}}    
    \includegraphics[width=0.7\linewidth]{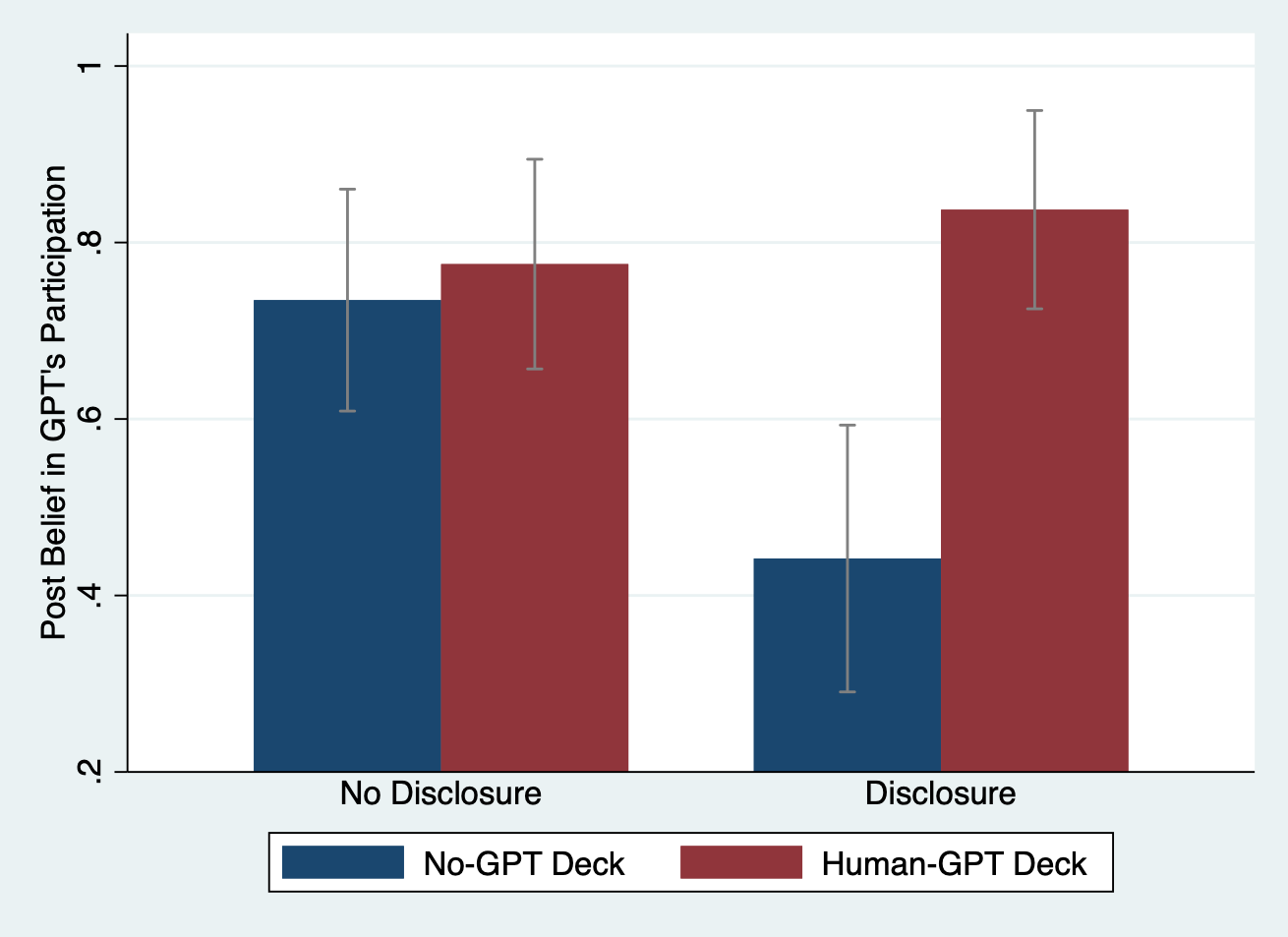}\\
    Note: The error bars indicate 95\% confidence intervals.
\end{figure}

We first test whether managers could identify the content created by analysts' use of GPT without disclosure. Figure \ref{fig:post_belief} presents the average percentage of managers who believe in GPT's involvement in content generation \textit{after} evaluating Human-GPT and No-GPT decks under each experimental condition. While there was no difference in managers' prior belief in GPT's participation before content evaluation, the results in Figure \ref{fig:post_belief} suggest that managers could not discern GPT's involvement in content generation when its source is not disclosed ($M_{Human-GPT} = 77.55\%$, $M_{No-GPT} = 73.47\%$, $p=0.68$ given a Chi-squared test of independence), providing evidence of information asymmetry regarding GPT's participation in content contribution without disclosure. Notably, analysts' non-disclosure of GPT use could lead managers, in general, to suspect their use of GPT, given that the lower bound of the 95\% confidence intervals is above 50\% for both the No-GPT and Human-GPT decks.

We find that enforcing disclosure is effective in reducing the information asymmetry of analysts' GPT use, as managers become better at identifying the content generated with GPT's participation when such information is disclosed ($M_{Human-GPT} = 83.72\%$, $M_{No-GPT} = 44.19\%$, $p<0.001$ given a Chi-squared test of independence). Interestingly, 44.19\% of the time, managers still think analysts used GPT despite full transparency and honesty about the lack of GPT use. This partially resulted from the managers' lack of trust in the analysts, given our experimental setting, where managers could not hand-pick their analysts to form a team. Nevertheless, it suggests a more significant reduction in information asymmetry if managers could trust analysts' honest disclosure of GPT use.

\subsection{The Impact of Disclosure Policy on the Alignment in GPT Adoption Between the Manager-Agent Dyad}

The information asymmetry between analysts and managers regarding GPT's participation in content generation without enforcement of disclosure policies suggests a potential moral hazard. Analysts may benefit from productivity gains achieved through GPT usage while exposing managers to risks associated with GPT involvement \citep{harris1979optimal,bosse2016agency}. On the other hand, managers would incur monitoring costs or need to offer incentives for the analysts to disclose their GPT use if they benefit from such information. In sum, information asymmetry in analysts' GPT usage, or its reduction, suggests a potential shift in the incentives for GPT adoption among the manager-analyst dyad. In this section, we investigate the incentives for managers and analysts to adopt GPT and how these align with variations in the disclosure policy.

We gauge analysts' incentives for adopting GPT by their managers' quality evaluations of the research briefs and the perceived effort exerted (in hours) in creating them. We generalize the evaluations (either perceived quality or exerted effort) from manager $i$ of deck version $v$ ($v\in\{No-GPT, Human-GPT\}$) for RFP case $c$ ($c\in\{1,2,3,4\}$) as $Y_{i,c,v}$. We quantify analysts' interest in adopting GPT in Equation \ref{eq:analyst} where the focal explanatory variable is $\text{Human-GPT}_{c,v}$ (coded as 1 for the Human-GPT deck and 0 for the No-GPT deck given an RFP $c$) and its interaction with $\text{Disclosure}_{i}$ (coded as 1 for the disclosure and 0 for the non-disclosure experimental condition). We include RFP case fixed effects $\eta_c$ to account for any observed and unobserved heterogeneity in the evaluation stemming from the time-invariant features across the four RFP cases. We also account for manager-level heterogeneity by including manager-level fixed effects $\gamma_i$ in the specification, which absorb variations of the $\text{Disclosure}_{i}$ main effect. According to this specification, a positive (negative) coefficient estimate of $\beta_1$ indicates that analysts would receive better (worse, respectively) evaluations and greater (less, respectively) perceived effort by the managers' assessment of the Human-GPT versus the No-GPT deck under the non-disclosure policy. It is important to recognize that, as managers cannot identify the actual content generation source, $\beta_1$ accurately reflects the changes in analysts' performance or effort as perceived by the managers when utilizing GPT for research briefs. $\beta_2$ measures the impact of the disclosure policy against the non-disclosure policy on changes in evaluation: a positive (negative) coefficient suggests an increase (decrease, respectively) in the evaluation gain or loss from analysts' use of GPT to create the decks under the disclosure policy relative to the non-disclosure policy. Finally, the combined effect of $\beta_1$ and $\beta_2$ reflects the increase or decrease in managers' evaluations of a Human-GPT deck relative to the No-GPT deck under the disclosure policy, suggesting greater or less incentive for the analyst to use GPT under an enforced disclosure policy.

\begin{align*}
Y_{i,c,v} & =\beta_0+\beta_1\times\text{Human-GPT}_{c,v}+\beta_2\times\text{Human-GPT}_{c,v}\times\text{Disclosure}_i+\eta_c+\gamma_{i}+ \epsilon_{i,c,v} \numberthis\label{eq:analyst}
\end{align*}



We measure manager $i$'s incentive to adopt GPT by their stated preference to authorize analysts to use GPT for each content section $s$ of a research brief $\text{SecGPTAuth}_{i,s}$ after evaluating the research briefs with or without the disclosure policy. Following the literature in the technology acceptance model \citep[e.g.,][]{agarwal1998conceptual,davis1989perceived,venkatesh2000theoretical}, we specify in Equation \ref{eq:manager} a manager $i$'s preference in authorizing GPT use influenced by the perceived usefulness ($\text{IndGPTUsefulness}_{i}$) and risk ($\text{IndGPTRisk}_{i}$) of GPT (both mean-centered) in contributing to the research brief. Managers' perceived risk of GPT use is measured by the average perception of misinformation raised from GPT responses and confidentiality breaches during analysts' interaction with GPT. We note that although managers perceived GPT as useful ($M_{Non-Disclosure}=5.09$, $M_{Disclosure}=5.07$, $p=0.91$ given a two-sample Student's \textit{t}-test) on a 7-point scale, managers state significant risk that GPT could potentially bring to content generation ($M_{Non-Disclosure}=4.58$, $M_{Disclosure}=4.56$, $p=0.82$ given a two-sample Student's \textit{t}-test) on a 5-point scale. We include research brief section-fixed effects $\gamma_s$ to account for any observed and unobserved heterogeneity in the tendency to authorize analysts' GPT use across the research brief content sections. We account for manager-level heterogeneity by including manager-level controls $X_i$, which compromise the manager's gender, working hours per week, whether their work faces clients, whether they evaluated the No-GPT before the Human-GPT deck, and tenure as a consultant. According to the specification in Equation \ref{eq:manager}, the coefficient estimates of $\gamma_1$ and $\gamma_2$ indicate how the perceived risk and usefulness influence managers' preference to adopt GPT under the non-disclosure policy. The coefficient estimate of $\gamma_3$ suggests managers' preference to adopt GPT under disclosure relative to the non-disclosure policy. Notably, the coefficient estimates of $\gamma_4$ and $\gamma_5$ measure the influence of the disclosure policy compared to the non-disclosure policy on shifts in managers' preference to adopt GPT driven by its perceived usefulness and associated risk. Consequently, the joint impact of $\gamma_1$, $\gamma_3$, and $\gamma_4$ shows how the perceived GPT risk affects managers' preference to adopt GPT under the disclosure policy; similarly, the combined influence of $\gamma_2$, $\gamma_3$, and $\gamma_5$ highlights how the perceived usefulness of GPT affects managers' preference to adopt it under the same disclosure policy.

\begin{align*}
SecGPTAuth_{i,s} & =\gamma_0+\gamma_1\times\text{IndGPTRisk}_{i}+\gamma_2\times\text{IndGPTUsefulness}_{i}+ \gamma_3\times\text{Disclosure}_{i}\\
& \gamma_4\times\text{IndGPTRisk}_{i}\times\text{Disclosure}_{i}+\gamma_5\times\text{IndGPTUsefulness}_{i}\times\text{Disclosure}_{i}+ \\
& \gamma_6\times X_i+\gamma_{s}+ \epsilon_{i,s} \numberthis\label{eq:manager}
\end{align*}

We consider managers and analysts share aligned incentives to adopt GPT when analysts are likely to receive better evaluations or be recognized for their greater effort by the managers in creating content for the Human-GPT deck compared to the No-GPT deck, while the managers would not forbid the analysts' GPT use. Importantly, we investigate how the alignment in the preference to adopt GPT varies when changing the disclosure policy. In particular, the consistent direction in the coefficient estimates for $\beta_2$, $\gamma_4$, and $\gamma_5$ implies a stronger alignment among the manager-analyst dyad in using GPT when enforcing a disclosure policy.

We apply the ordinary least square (OLS) estimation with robust standard errors clustered at the individual manager level for all analyses to allow for arbitrary error correlations within each manager. Table \ref{tab:dilemma_moderation} presents the results. In particular, Columns (1) and (2) show the results related to analysts' incentive to adopt GPT by using managers' deck evaluation and effort estimation as the dependent variable following the specification shown in Equation \ref{eq:analyst}; Column (3) shows the results about managers' preference to adopt GPT according to the specification in Equation \ref{eq:manager}. In the following sections, we interpret the results separately under the non-disclosure and disclosure conditions.

\subsubsection{Non-Disclosure Policy}

We first investigate analysts' and managers' incentives in adopting GPT under the non-disclosure condition, the status quo, where the content generation source is obscured from the managers. 

For the analysts, the positive and statistically significant coefficient estimate of $\beta_1$ in Table \ref{tab:dilemma_moderation} Column (1) indicates that analysts benefit from using GPT to produce a research brief, as they receive better-quality evaluations from their superiors without disclosing its use. However, based on the coefficient estimate of $\beta_1$ in Table \ref{tab:dilemma_moderation} Column (2), we do not find that managers consider generating this better content to require more significant effort from the analysts.

For the managers, the positive and statistically significant coefficient estimate of $\gamma_2$ in Table \ref{tab:dilemma_moderation} Column (3) suggests that higher perceived usefulness of GPT increases managers' preference to authorize analysts' GPT use. However, the negative and statistically significant coefficient estimate of $\gamma_1$ in Table \ref{tab:dilemma_moderation} Column (3) indicates that a higher perceived risk stemming from using GPT decreases managers' incentives to authorize the use of GPT. Since $IndGPTRisk$ and $IndGPTUsefulness$ are mean-centered, managers' average tendency to adopt GPT is 2.309 (coefficient estimate of $\gamma_0$), which is marginally statistically significant below the mid-point indicating adoption difference of the 5-point scale ($p=0.05$ given a Z-test). Although managers considering GPT to be more useful (one standard deviation above the sample mean) do not show a significant preference for forbidding analysts' GPT use ($p=0.97$ given a Z-test compared to the mid-point indicating adopting indifference), managers who hold greater risk concerns (one standard deviation above the sample mean) strongly forbid analysts' GPT use ($p<0.01$ given a Z-test compared to the mid-point indicating adopting indifference).

The findings suggest that analysts could benefit from using GPT without disclosing it, as this could lead to better performance evaluations from their superiors. However, managers do not show a general positive tendency to adopt GPT, unless they consider GPT beneficial or at less risk of generating misinformation or confidentiality breaches when they do not know whether the analysts used GPT. Although a non-disclosure policy could encourage analysts' adoption, their superior would not authorize its use at large, leading to misaligned incentives in adopting GPT among the manager-analyst dyad. These findings corroborate the existing literature showing that GPT adoption frictions are primarily caused by the reservations of superiors \citep{humlum2024adoption} and explain why subordinates exhibit shadow adoption of GPT when enterprise-wide adoption is not implemented.

\begin{table}[!h]
    \centering
    \caption{The Impact of Disclosure on Manager-Analyst Incentives to Adopt GPT\label{tab:dilemma_moderation}}
    \begin{threeparttable}
        \resizebox{1.05\textwidth}{!}{%
        {
\def\sym#1{\ifmmode^{#1}\else\(^{#1}\)\fi}
\begin{tabular}{l*{3}{c}}
\hline\hline
                         &          DV: DeckQuality   &     DV: $\log$(DeckHour)   &           DV: SecGPTAuth   \\
                         &  \multicolumn{1}{c}{(1)}   &  \multicolumn{1}{c}{(2)}   &  \multicolumn{1}{c}{(3)}   \\
\hline
Human-GPT (vs. No-GPT)   &                    0.389***&                    0.079   &                            \\
                         &                  (0.127)   &                  (0.053)   &                            \\
[1em]
Human-GPT (vs. No-GPT) $\times$ Disclosure&                   -0.219   &                   -0.387***&                            \\
                         &                  (0.182)   &                  (0.091)   &                            \\
[1em]
IndGPTRisk               &                            &                            &                   -0.514***\\
                         &                            &                            &                  (0.156)   \\
[1em]
IndGPTUsefulness         &                            &                            &                    0.517***\\
                         &                            &                            &                  (0.107)   \\
[1em]
IndGPTRisk $\times$ Disclosure&                            &                            &                    0.549** \\
                         &                            &                            &                  (0.273)   \\
[1em]
IndGPTUsefulness $\times$ Disclosure&                            &                            &                   -0.226   \\
                         &                            &                            &                  (0.137)   \\
[1em]
Disclosure               &                            &                            &                    0.102   \\
                         &                            &                            &                  (0.153)   \\
[1em]
Constant                 &                    3.992***&                    2.286***&                    2.309***\\
                         &                  (0.046)   &                  (0.023)   &                  (0.354)   \\
\hline
Observations             &                      238   &                      238   &                      552   \\
Individual fixed-effects &                      Yes   &                      Yes   &                       No   \\
Case fixed-effects       &                      Yes   &                      Yes   &                       No   \\
Pitch section fixed-effects&                       No   &                       No   &                      Yes   \\
Individual controls      &                       No   &                       No   &                      Yes   \\
\hline\hline
\end{tabular}
}
}
        \begin{tablenotes}
            \item\parbox{\textwidth}{Notes: Standard errors in the parentheses are clustered at the individual manager level. * $p<0.10$, ** $p<0.05$, *** $p<0.01$. }
        \end{tablenotes}
    \end{threeparttable}
\end{table}

\subsubsection{Disclosure Policy}

We next investigate the incentives of analysts and managers adopting GPT under the disclosure condition where the content generation source is communicated to the managers. 


The results shown in Table \ref{tab:dilemma_moderation} Columns (1) suggest that analysts do not necessarily receive significantly worse evaluations from the managers when disclosing the GPT use for the Human-GPT deck given the non-statistically significant coefficient estimate of $\beta_2$. However, the advantage of the better-evaluated Human-GPT versus the No-GPT decks disappeared, as
the combined coefficient estimate of $\beta_1$ and $\beta_2$ is not statistically different from zero ($p=0.17$ given a Z-test). Notably, the negative statistically significant coefficient estimate of $\beta_2$ shown in Table \ref{tab:dilemma_moderation} Column (2) suggests that managers perceive analysts to put in less effort in creating the Human-GPT deck when GPT use is disclosed compared to the No-GPT deck.

For the managers, the results shown in Table \ref{tab:dilemma_moderation} Column (3) suggest: 1) the disclosure policy does not positively encourage managers to adopt GPT given the non-significant coefficient estimate of $\gamma_3$, and 
2) the positive and statistically significant coefficient estimates of $\gamma_4$ suggest that an enforced disclosure policy could reduce the negative impact of risk concerns on managers' preferences to authorize the GPT use under the non-disclosure policy. In particular, managers with greater risk concerns (one standard deviation above the sample mean) do not show a significant preference for forbidding analysts' GPT use ($p=0.14$ given a Z-test compared to the mid-point indicating adopting indifference). Importantly, our findings indicate that the enforced disclosure policy does not necessarily alleviate the broader misaligned interests between managers and analysts in utilizing GPT under a non-disclosure agreement. This conclusion is supported by the inconsistent signs in the coefficient estimates of $\beta_2$ and $\gamma_4$, as presented in Table \ref{tab:dilemma_moderation}, Columns (2) and (3).

In conclusion, the findings indicate that analysts might forfeit the advantages of GPT in enhancing content evaluation if they reveal their use to managers. In particular, the performance gain from using GPT has vanished compared to instances of non-disclosure, accompanied by less effort attributed to their content generation. On the other hand, our results suggest that an enforced disclosure policy could reduce the negative impact of managers' concerns about GPT risks on their incentive to adopt GPT when no such policy exists. This finding underscores the conflicting incentives for managers and analysts regarding GPT adoption, which a disclosure policy does not resolve.

\subsection{Discussion}

Our survey experiment results suggest that GPT has the potential to enhance content quality for professional service firms, aligning with existing research \citep[e.g.,][]{Noy2023-se}. However, the successful integration of GPT in the workplace depends on aligned incentives for adoption and accountability in risk management, as seen through the lens of agency theory. To our knowledge, we are the first to demonstrate empirically that achieving such alignment poses significant challenges in a professional service firm, which adds to the well-documented friction surrounding GPT adoption across industries \citep{humlum2024adoption}.

Professional service firms without an explicit GPT adoption and reinforced disclosure policy are subject to potential agency costs due to information asymmetry that managers cannot tell analysts' GPT use. In this case, firms cannot harness the benefits of analysts adopting GPT in content generation due to managers' hesitance to allow its use stemming from risk concerns about GPT's contribution to misinformation and confidentiality breaches. As a result, the information asymmetry in analysts' GPT uses implies a potential increase in monitoring costs and exacerbates the incentive misalignment in the manager-analyst dyad in enterprise-wide GPT adoption.

Although our results suggest an enforced disclosure policy would reduce the information asymmetry that causes the potential monitoring cost mentioned above, this policy alone would not reduce the misalignment in the incentives between managers and analysts in adopting GPT. On the one hand, the reinforcement in the disclosure policy could encourage managers' incentives to adopt GPT by reducing the negative impact of the risk concerns. On the other hand, the enforced disclosure policy makes analysts less likely to use GPT because managers tend to undervalue the performance and effort put in by analysts when knowing their GPT use. In professional service organizations, when the effort is undervalued, it suggests that managers have an inflated view of analysts' capacities, resulting in an anticipated reduction in analysts' hourly wages without improved performance assessments. Furthermore, the misaligned incentives in GPT adoption between managers and agents could be further worsened if analysts are also made accountable for any misinformation or confidentiality breaches when they disclose its use. The implications suggest that implementing a disclosure policy should be further accompanied by corporate AI policies that address accountability in risk-sharing and updated salary schemes related to GPT integration in the workplace.

\section{Potential Solutions to Better Align Incentives of the Manager-Analyst Dyad in GPT Adoption}

We have so far identified a general trend of misaligned incentives regarding the adoption of GPT among the manager-agent dyad, which points to a broader dilemma in GPT adoption with and without a disclosure policy. However, our empirical investigation does not fully explore whether some managers share a more aligned preference with an analyst in utilizing GPT. This could offer a preliminary solution and recommendation for a corporate AI policy, which we will discuss in this section.

\subsection{Impact of Managers' Acknowledgment on Human Effort in the Alignment of GPT Adoption Among the Manager-Analyst Dyad}

The findings from the previous section indicate that misalignment incentives between managers and analysts regarding the use of GPT occur in two ways: first, in situations lacking a disclosure policy, managers are reluctant to embrace GPT due to risk apprehensions. Second, with a disclosure policy in place, managers often underestimate analysts' performance and the effort required to prepare research briefs involving GPT. These situations highlight that for GPT to be effectively integrated into the manager-analyst dynamic, managers should recognize the human effort in content creation and allow the use of GPT, even amid significant risk concerns. Consequently, we focus on a specific group of managers who value analysts' efforts when utilizing GPT. This group usually possesses a more realistic view of their employees' abilities, which aids in maintaining analysts' hourly wages by controlling the number of tasks assigned in a given period. While safeguarding analysts' income, we also evaluate whether these managers have a favorable opinion of their subordinates' performance. Additionally, we explore if implementing a disclosure policy could help mitigate the negative impacts of risk on allowing analysts' GPT use.

We consider managers who would maintain their subordinates' workload when utilizing GPT if the relative estimated number of hours used to generate the Human-GPT is not less than that of the No-GPT deck for the same RFP case. Furthermore, we do not view managers as realistic if they consider a deck that requires less than four hours of work since the actual time spent generating a deck is approximately four hours without GPT. Table \ref{tab:effort} Column (1) shows the subsample results related to analysts' incentive to adopt GPT by using managers' deck evaluation using the exact specification as shown in Table \ref{tab:dilemma_moderation} Column (1).\footnote{We do not investigate analysts' incentive to use GPT by managers' effort estimation since managers in this subgroup do not discount analysts' efforts in developing the decks.} Table \ref{tab:effort} Column (2) shows the results regarding the same subgroup of managers' preference to adopt GPT according to the exact specification as in Table \ref{tab:dilemma_moderation} Column (3).

We first investigate analysts' and managers' incentives for adopting GPT under the non-disclosure policy. For the analysts, the positive and statistically significant coefficient estimate of $\beta_1$ in Table \ref{tab:effort} Column (1) indicates that analysts benefit from using GPT to produce a research brief. For the managers, their average tendency to adopt GPT is 2.569 (coefficient estimate of $\gamma_0$), which is not statistically significant below the mid-point indicating adoption difference of the 5-point scale ($p=0.44$ given a Z-test), suggesting no significant preference to forbid analysts' GPT use. Consistent with the results shown in Table \ref{tab:dilemma_moderation} Column (3), the positive and statistically significant coefficient estimate of $\gamma_2$ in Table \ref{tab:effort} Column (2) suggests that a higher perceived usefulness of GPT increases managers' preference to authorize analysts' GPT use. The negative and statistically significant coefficient estimate of $\gamma_1$ in Table \ref{tab:effort} Column (2) indicates that a higher perceived risk stemming from the use of GPT decreases managers' incentives to authorize the use of GPT. Managers contemplating the risks associated with GPT, defined as one standard deviation above the average, exhibit a marginally statistically significant tendency to prohibit analysts from using GPT (p=0.10, based on a Z-test compared to the mid-point that suggests a stance of indifference). In summary, managers who acknowledge the importance of human effort in creating the Human-GPT, as opposed to the No-GPT deck without a disclosure policy, not only see improved performance but also would not forbid GPT usage unless they identify significant risks. 

We next investigate analysts' and managers' incentives for adopting GPT under the disclosure policy. The findings in Table \ref{tab:effort}, Columns (1), indicate that analysts do not appear to be evaluated significantly worse by managers when revealing GPT use for the Human-GPT deck, as evidenced by the non-statistically significant coefficient estimate of $\beta_2$. However, the advantage of the better-evaluated Human-GPT compared to the No-GPT decks disappeared, as the combined coefficient estimate of $\beta_1$ and $\beta_2$ is not statistically different from zero ($p=0.85$ based on a Z-test). For managers, the outcomes displayed in Table \ref{tab:effort} Column (2) indicate that implementing a disclosure policy might lessen the adverse effect of risk concerns on their willingness to approve GPT usage under the non-disclosure policy, as evidenced by the positive and statistically significant coefficient estimates of $\gamma_4$. Specifically, managers with heightened risk concerns (one standard deviation above the sample mean) exhibit no significant preference against prohibiting analysts' use of GPT ($p=0.74$ following a Z-test compared to the mid-point, suggesting a stance of indifference). Consequently, even though managers permitted their subordinates to utilize GPT under the disclosure policy, the previously recognized enhancement in performance of the Human-GPT compared to the No-GPT deck under the non-disclosure policy vanished. This suggests a misalignment of incentives within the manager-analysts relationship regarding the GPT adoption, despite managers not undermining analysts' efforts.  

\begin{table}[htbp]
    \centering
    \caption{The Impact of Disclosure on Manager-Analyst Incentives to Adopt GPT–Subsample Analysis When Managers Do Not Discount Effort When Analysts Used GPT \label{tab:effort}}
    \begin{threeparttable}
        \resizebox{.8\textwidth}{!}{%
        {
\def\sym#1{\ifmmode^{#1}\else\(^{#1}\)\fi}
\begin{tabular}{l*{2}{c}}
\hline\hline
                         &              DV: Quality   &           DV: SecGPTAuth   \\
                         &  \multicolumn{1}{c}{(1)}   &  \multicolumn{1}{c}{(2)}   \\
\hline
Human-GPT (vs. No-GPT)   &                    0.437***&                            \\
                         &                  (0.156)   &                            \\
[1em]
Human-GPT (vs. No-GPT) $\times$ Disclosure&                   -0.403   &                            \\
                         &                  (0.243)   &                            \\
[1em]
IndGPTRisk               &                            &                   -0.772***\\
                         &                            &                  (0.210)   \\
[1em]
IndGPTUsefulness         &                            &                    0.578***\\
                         &                            &                  (0.151)   \\
[1em]
IndGPTRisk $\times$ Disclosure&                            &                    1.056***\\
                         &                            &                  (0.339)   \\
[1em]
IndGPTUsefulness $\times$ Disclosure&                            &                   -0.209   \\
                         &                            &                  (0.197)   \\
[1em]
Disclosure               &                            &                    0.440** \\
                         &                            &                  (0.209)   \\
[1em]
Constant                 &                    4.213***&                    2.569***\\
                         &                  (0.060)   &                  (0.558)   \\
\hline
Observations             &                      146   &                      246   \\
Individual fixed-effects &                      Yes   &                       No   \\
Case fixed-effects       &                      Yes   &                       No   \\
Pitch section fixed-effects&                       No   &                      Yes   \\
Individual controls      &                       No   &                      Yes   \\
\hline\hline
\end{tabular}
}
}
        \begin{tablenotes}
            \footnotesize
            \item\parbox{\textwidth}{Notes: Standard errors in the parentheses are clustered at the individual manager level. * $p<0.10$, ** $p<0.05$, *** $p<0.01$. }
        \end{tablenotes}
    \end{threeparttable}
    
\end{table}

\subsection{Impact of Managers' Experience on the Alignment in GPT Adoption Among the Manager-Analyst Dyad Under the Disclosure Policy}

Managers and their analysts show an alignment in incentives to use GPT within the non-disclosure policy when the managers acknowledge the effort put in by analysts in creating the Human-GPT deck. However, this optimism is still undermined by the disclosure policy, largely influenced by managers' diminished assessment of content generation quality due to analysts' use of GPT. This leads us to investigate whether differences exist among managers in their evaluations of the Human-GPT versus No-GPT presentations and their preference to allow GPT utilization. Previous studies indicate that managers' experience influences their willingness to embrace AI and its output \citep{logg2019algorithm,Noy2023-se,peng2023impact,DellAcqua2023-mj}. We aim to empirically assess how the incentives for adopting GPT align between managers and analysts based on the managers' experience level under the disclosure policy. We concentrate on the disclosure condition of the subsample analyzed in Table \ref{tab:effort} and exclude observations from the non-disclosure condition. Keep in mind that excluding these non-disclosure observations alters the specification, as indicated by Equations \ref{eq:analyst} and \ref{eq:manager}, since it removes terms associated with $Disclosure$. 

To measure analysts' incentive to use GPT, we incorporate managers' experience through an interaction term between $\text{IndExperience}$ (mean-centered) and $\text{Human-GPT}$. This enables us to examine how variations in managers' experience affect their assessments of content quality in the decks prepared by the analysts.\footnote{The primary influence of $\text{IndExperience}$ is captured by the individual-level fixed effects.} The OLS regression results are shown in Table \ref{tab:experience} Column (1). Although managers, on average, do not recognize the improved quality of the Human-GPT (vs. No-GPT) deck, more experienced managers significantly value the Human-GPT deck compared to less experienced managers, given the positive and statistically significant coefficient estimate of the interaction term between $\text{Human-GPT}$ and $\text{IndExperience}$. Moreover, more experienced managers (one standard deviation above the mean) significantly prefer the Human-GPT deck over the No-GPT deck ($p<0.05$ based on a Z-test). In contrast, this favorable assessment is absent among less experienced managers ($p=0.12$ according to a Z-test).

To measure managers' incentive to adopt GPT, we introduce $\text{IndExperience}$ and its interaction with $\text{IndGPTRisk}$ and $\text{IndGPTUsefulness}$ to understand whether managers' experience would change their tendency to allow GPT use. The results are shown in Table \ref{tab:experience} Column (2). On average, managers do not exhibit a notable preference against allowing analysts to use GPT. This conclusion stems from the constant term's coefficient estimate of 3.171, which is not statistically significant under the midpoint of the 5-point scale ($p = 0.79$ from a Z-test). While more experienced managers show greater awareness of the risks associated with adopting GPT for analysts, they also perceive greater benefits in using GPT compared to their less experienced counterparts. We observe that managers, regardless of their experience level, do not significantly favor banning analysts' GPT ($ps > 0.85$ as indicated by Z-tests to the preference midpoint, implying a neutral position).

In conclusion, managers with more experience who value analysts' contributions in utilizing GPT for content creation help reduce the misaligned incentives in GPT adoption between the manager-agent relationship under the disclosure policy.

\begin{table}[hbtp]
    \centering
    \small
    \caption{The Impact of Managers' Experience on Manager-Analyst Incentives to Adopt GPT under the Disclosure Policy\label{tab:experience}}
    \begin{threeparttable}
        {
\def\sym#1{\ifmmode^{#1}\else\(^{#1}\)\fi}
\begin{tabular}{l*{2}{c}}
\hline\hline
                         &              DV: Quality   &           DV: SecGPTAuth   \\
                         &  \multicolumn{1}{c}{(1)}   &  \multicolumn{1}{c}{(2)}   \\
\hline
Human-GPT (vs. No-GPT)   &                    0.013   &                            \\
                         &                  (0.156)   &                            \\
[1em]
Human-GPT (vs. No-GPT) $\times$ IndExperience&                    0.362** &                            \\
                         &                  (0.148)   &                            \\
[1em]
IndGPTRisk               &                            &                   -0.563   \\
                         &                            &                  (0.376)   \\
[1em]
IndGPTUsefulness         &                            &                    0.235   \\
                         &                            &                  (0.214)   \\
[1em]
IndGPTRisk $\times$ IndExperience&                            &                   -0.984** \\
                         &                            &                  (0.394)   \\
[1em]
IndGPTUsefulness $\times$ IndExperience&                            &                    0.202** \\
                         &                            &                  (0.094)   \\
[1em]
IndExperience            &                            &                    0.490   \\
                         &                            &                  (0.385)   \\
[1em]
Constant                 &                    4.322***&                    3.171***\\
                         &                  (0.077)   &                  (0.630)   \\
\hline
Observations             &                       58   &                      114   \\
Individual fixed-effects &                      Yes   &                       No   \\
Case fixed-effects       &                      Yes   &                       No   \\
Pitch section fixed-effects&                       No   &                      Yes   \\
Individual controls      &                       No   &                      Yes   \\
\hline\hline
\end{tabular}
}

        \begin{tablenotes}
            \footnotesize
            \item Notes: Standard errors in the parentheses are clustered at the individual manager level. * $p<0.10$, ** $p<0.05$, *** $p<0.01$. 
        \end{tablenotes}
    \end{threeparttable}
\end{table}

\subsection{Discussion}

Although the primary findings from section \ref{main_result} indicate a misalignment between managers and analysts in adopting GPT, this section explores specific groups of managers who improve alignment within the manager-analyst relationship regarding GPT adoption. Specifically, our findings indicate that managers who do not diminish the significance of human effort exerted in content generation could perceive that analysts perform better when GPT is employed, depending on either a non-disclosure policy or a disclosure policy restricted to more experienced managers. These findings indicate that professional service firms should focus on identifying experienced managers who do not overlook the efforts of their subordinates when using GPT. This method proves advantageous when corporate AI policies are lacking. Specifically, a company can gain productivity through GPT integration while concurrently developing and enacting AI policies that clarify risk-sharing responsibilities and introduce revised salary frameworks associated with GPT implementation in the workplace.

Our findings advance the emerging literature on GPT adoption 
\citep[e.g.,][]{Noy2023-se,brynjolfsson2023generative,DellAcqua2023-mj}. While prior studies indicate that less experienced consulting workers might have greater incentives to adopt GPT 
\citep{DellAcqua2023-mj}, our results imply that this adoption strategy may be inefficient in a manager-analyst dynamic, especially if consulting workers hold managerial roles. We also highlight the significance of experienced managers within consulting firms. Although they may not gain the most immediate benefits from directly using GPT, they tend to support their subordinates' use of GPT when they recognize its advantages. Furthermore, we observe that the disclosure policy fosters this alignment of interests while managers still account for the hours analysts spend creating the deck using GPT, pointing to a need for policies that ensure managers objectively acknowledge analysts' contributions when leveraging GPT.

\section{Conclusions}

The empirical results highlight novel problems associated with information asymmetry and moral hazard that may arise from subordinates' adoption of ChatGPT in consulting firms. We specifically identified how these agency problems affect the adoption of ChatGPT and the management of risks associated with misinformation and confidentiality. Agency scholarship identifies multiple solutions to address agency problems in firms, including the adoption of monitoring mechanisms and incentive structures \citep{eisenhardt1989agency}, as well as the implementation of cooperation strategies between individuals and teams \citep{miller2005solutions}. While specific actions based on these solutions are certainly needed, the question remains regarding the appropriate instruments to implement and formalize this wide array of solutions \citep{brudney1996contract}.

We suggest that the adoption of AI corporate policies is the most effective means to address these agency problems for multiple reasons. First, the absence of a corporate-wide policy would result in agency problems arising from the individual use of ChatGPT being handled on a case-by-case basis, exposing the firm to multiple and diverse risks associated with each single solution and adding friction to ChatGPT adoption. Second, the implementation of an AI corporate policy introduces a contractual component within the firm, whereby the duties between principals and agents related to the use of ChatGPT and other generative AI software become part of the ``agency law within the corporation." This means that agents within the firm will be explicitly bound through a contractual duty to the principal, aligning both parties in terms of information sharing and responsibilities. Finally, an AI corporate policy ensures that risks are shared and known to both principals and agents, aligning the adoption of ChatGPT not only with the interests of managers but also with the broader interests of the firm.

We propose that an AI corporate policy should encompass at least four key elements to effectively address agency problems resulting from the agents' use of ChatGPT. First, the policy should mandate a disclosure obligation requiring agents to inform principals whenever they use ChatGPT for task completion. This transparency reduces information asymmetry, enabling managers to request and access the necessary information to assess the quality and authenticity of the work produced, and implement appropriate controls to mitigate risks associated with misinformation and confidentiality breaches.  Second, the policy should establish a risk-sharing framework to align the interests of both principals and agents. By formally incorporating the use of ChatGPT or similar tools into the contractual duties of employees, both principals and agents are aware of their shared responsibilities and accountabilities, addressing moral hazard by ensuring agents are not incentivized to take undue risks, such as improperly sharing confidential information or over-relying on AI-generated content without oversight. Third, a monitoring mechanism should be implemented to ensure agents respect their duties. This mechanism should allow for the verification of whether agents disclose ChatGPT use, assess the types of risks agents incur by using ChatGPT, and review the actual outputs produced with ChatGPT to avoid legal issues such as misrepresentation and copyright violations. Finally, an incentive mechanism should be implemented to acknowledge the work done by agents with the help of ChatGPT without discounting their effort. This mechanism is necessary to align the interests of agents using ChatGPT with those of the principals in serving clients effectively.

This approach to corporate AI policy includes tracking ChatGPT usage, which would incur agency costs for managers. However, it embodies the intra-corporate duty of candor, as advocated by Langevoort, which imposes an affirmative duty on employees to disclose relevant information. Furthermore, we believe that monitoring ChatGPT use is essential to prevent diminished work quality and inadequate risk management. The Hamburg Labor Court case illustrates that simply authorizing the use of ChatGPT does not constitute a comprehensive policy, nor does it grant employers oversight of employee usage. Despite having AI guidelines and a handbook, the company remained unaware of ChatGPT adoption, revealing a critical gap in management’s visibility. These findings underscore the importance of disclosure, aligning with regulatory frameworks such as the AI Act (Regulation (EU) 2024/1689), which requires providers and deployers of AI systems to ensure their staff and other individuals involved in the operation and use of AI systems have sufficient AI literacy (Article 4). This obligation came into effect on February 2, 2025. At present, it remains unclear whether employers will be considered ``deployers" when they simply allow employees to use freely accessible AI systems, such as ChatGPT. However, this qualification is certainly possible, particularly if employers implement specific GPT use policies. Furthermore, employers could also be considered ``providers" of AI systems if they customize GPT-based AI systems for internal use. Consequently, we believe that employers, as part of their AI literacy initiatives, should pay special attention to disclosure and the responsible use of AI.
Consistent with existing literature, our analysis reaffirms that disclosure is an effective mechanism for addressing misalignment of interests in the workplace. The rise of generative AI presents a timely opportunity to strengthen this principle.

\singlespacing

\newpage
\printbibliography

\newpage
\section*{Appendix}
\setcounter{table}{0}
\setcounter{figure}{0}
\renewcommand{\thetable}{A\arabic{table}}
\renewcommand{\thefigure}{A\arabic{figure}}

\begin{figure}[!h]
    \centering
    \caption{Number of prompts used in each content generation section\label{fig:num_prompts}}
    \includegraphics[width=0.6\linewidth]{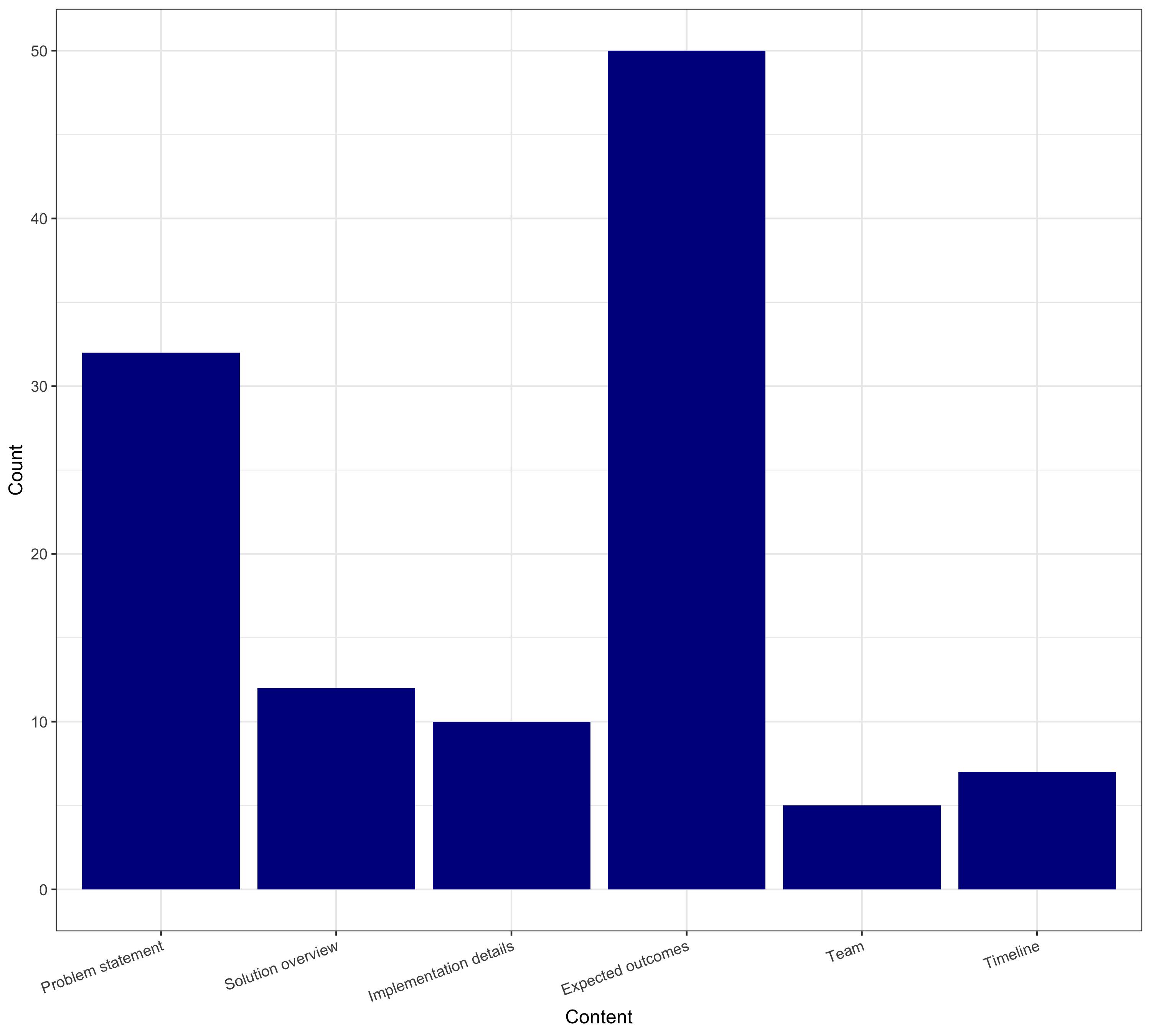}
\end{figure}

\begin{table}[!h]
    \centering
    \caption{Items used to measure the constructs\label{tab:item}}
    {
\def\sym#1{\ifmmode^{#1}\else\(^{#1}\)\fi}
\begin{tabular}{p{10em}p{26em}}
                        \hline\hline\\[-2ex] 
                        Construct & Item \\\hline
                        DeckQuality & 1. The problem description is accurate and clear.\\
                        & 2. The content is highly informative.\\
                        & 3. The proposed solutions are feasible.\\
                        \hline
                        IndGPTUsefulness & 1. How useful is ChatGPT for generating a research brief given the summarizing prompts? \\
                        & 2. How useful is ChatGPT for generating a research brief given the inferring prompts? \\
                        & 3. How useful is ChatGPT for generating a research brief given the text transforming prompts?\\
                        & 4. How useful is ChatGPT for generating a research brief given the expanding prompts? \\
                        \hline
                        IndGPTRisk & 1. Do you worry about misinformation generated by ChatGPT for your work?\\
                        & 2. Are you concerned about confidentiality issues related to content input into ChatGPT for your work?\\
                        \hline
                        
                \end{tabular}
}
\end{table}

\begin{table}[!h]
    \centering
    \caption{Variable name and definition\label{tab:varname}}
    {
\def\sym#1{\ifmmode^{#1}\else\(^{#1}\)\fi}
\begin{tabular}{p{10em}p{31em}}
                        \hline\hline\\[-2ex] 
                        Variable Name & Explanation \\\hline
                        \multicolumn{2}{p{41em}}{\textit{Individual-level Variables}}\\
                        IndGPTUsefulness & An individual manager's perceived usefulness of GPT in generating a research brief. This index is the average of the four items shown in Table \ref{tab:item}.\\
                        IndGPTRisk & An individual manager's perceived riskiness of GPT in generating a research brief. 
                        This index is the average of the two items shown in Table \ref{tab:item}.\\
                        IndFemale & Whether an individual manager is female.\\ 
                        IndWeekWorkingHour & The average number of hours an individual manager worked per week (level 1: less than 45 hours; level 2: less than 50 hours; level 3: less than 60 hours; level 4: other).\\ 
                        IndTenure & The tenure of an individual manager in consulting (level 1: less than a year; level 2: one to five years; level 3: five to seven years; level 4: more than 7 years).\\ 
                        IndNoGPTDeckFirst & Whether an individual manager evaluated a No-GPT deck before evaluating a Human-GPT deck. \\
                        IndClientFacing & Whether an individual manager has a client-facing role. \\ 
                        IndExperience &  The familiarity with the RFP process on a scale from one to five. We take the average of the familiarity of the RFP process across the two evaluated cases for managers recruited from the first wave of data collection.\\
                        
                        
                        \multicolumn{2}{p{41em}}{\textit{Individual-case-deck-level Variables}}\\
                        
                        DeckQuality & An individual manager's perceived deck quality (No-GPT or human-GPT) given a case. The index is an average of four items shown in Table \ref{tab:item}.\\
                        DeckHour & The perceived number of hours used to generate a deck (No-GPT or human-GPT) given a certain case. \\
                        
                        \multicolumn{2}{p{41em}}{\textit{Individual-case section-level Variables}}\\
                        SecGPTAuth & The likelihood of an individual manager allowing the analysts to use GPT to prepare business content on a scale from one to five.\\
                        
                        \hline
                \end{tabular}
}
\end{table}


        

\end{document}